# Contribution of High Energy Physics Techniques to the Medical Imaging Field

Pierre-Etienne Vert,[a*] Jacques Lecoq,[a] Gerard Montarou,[a] Nicoleta Pauna,[a] Baptiste Joly,[a] Madjid Boutemeur,[b] Hervé Mathez,[b] Renaud Gaglione,[b] Patrick Le Dû[c]

[a]*Laboratoire de Physique Corpusculaire, 24 avenue des Landais, 63177 Aubière Cedex, France*

[b]*Institut de Physique Nucléaire de Lyon, Domaine scientifique de la Doua, 4 rue Enrico Fermi, 69622 Villeurbanne cedex, France*

[c]*CEA/DAPNIA, 91191 Gif sur Yvette, France*

**Abstract**

The purpose of this study was to show how advanced concepts of compact, lossless and "Time Of Flight" (TOF) capable electronics similar to those foreseen for the LHC and ILC experiments could be fairly and easily transferred to the medical imaging field through Positron Emission Tomography (PET). As a wish of explanation, the two overriding weaknesses of PET camera readout electronics, namely dead-time and timing resolution, were investigated analytically and with a Monte-Carlo simulator presently dedicated to this task. Results have shown there was room left for count rate enhancement through a huge decrease of the timing resolution well below the nanosecond. The novel electronics scheme suggested for PET in this paper has been partly inspired by the long experience led in High Energy Physics where the latter requirement is compulsory. Its structure entirely pipelined combined to a pixelation of the whole detector should allow dead-times to be suppressed, while the absence of devoted timing channel would remove the preponderant contributions to the timing resolution. To the common solution for timing would substitute an optimal filtering method witch clearly appears as a good candidate as timing resolution of a few tens of picoseconds may be achieved provided the shape of the signal is known and sufficient samples are available with enough accuracy. First investigations have yield encouraging results as a sampling frequency of 50 MHz with a 7 bits precision appears sufficient to ensure the 500ps coincidence timing resolution planed. At this point, there will be a baby step ahead to draw benefice from a TOF implementation to the design and the enormous noise variance enhancement that would come with.

PET, dead-time, NECR, timing resolution, electronics, TOF

———

* Corresponding author. Tel.: +33 (0)47-340-5146; fax: +33(0)47-326-4598; e-mail: vert@clermont.in2p3.fr.

## 1. Introduction

In 3D PET, the selection of events naturally applies in coincidence and is performed by the electronics alone. A coincidence window is hardware implemented and serves as a tolerance for the two photons to hit any part of the detector wherever lies the annihilation in the patient. Scattered and random events may arise in time and be accepted by the electronics, whereas they effectively act as noise. Besides occupying the acquisition chain and saturating it sometimes due to the dead-times shared along, the impeding presence of noise may appear as magnified at saturation as true events can be missed. Overall, it clearly contributes to alter the instrument sensitivity. The motivations behind our study was to evaluate the limitations in term of sensitivity of "recent" scanners and the potential improvement left when going to an architecture with performances pushed to their limits. In a first time, the results that outcome from a Monte-Carlo program specifically written to undertake the physics of emission in PET were re-injected into a simple behavioural model of a typical detection chain to compute intermediate and overall dead-times effect on count-rates as a function of the input entries. In a second time, we went through the timing resolution aspect and refer for that to a published investigation of the respective contributions of the components that make up the system regarded [1]. With the program, we checked the influence of this factor on count-rate performances when lowered to its physical limit by reflecting it to the coincidence window width. Finally, we merged both ideas and figured out the net gain that could be drawn from a lossless and high timing resolution electronics before suggesting the synoptic of what is actually being designed. The minimisation of the timing resolution opens the opportunity for the TOF to be re-introduced in PET as it used to in the 1980s. Indeed, the new heavy scintillator crystals like LSO or LaBr$_3$ now offer possibilities in terms of sensitivity and spatial resolution that were unattainable in the early years and prevent from major tradeoffs to be done [2]. For our approach to be global, we figured out the subsequent gain in noise variance that a TOF implementation could bring.

In the remainder of this paper, we will present a study that sticks to the HR+ PET camera from CTI. A simple reason to that is that most of the literature we found with sufficient details about the architecture covered this scanner in particular [3] [4]. We paid a special attention to make a comparison with an up-to-date instrument, the Accel from CPS. Both camera belong to the same family, as Siemens subsidiaries, and share parent ring hardware properties. Literature indicates they were developed with common electronics philosophy [1] [3]. Hence, replacing initial BGO scintillator of the HR+ by LSO to match the Accel on this point too, we re-adjusted parameters of the dead-time model so that the new camera, yet hypothetical, finally yields count rate performances close to the Accel [5].

## 2. Count rate performances

As mentioned in introduction, a simulator was specifically developed to run the physics of emission relative to the HR+ PET camera, along with the behaviour of the incoming events through the acquisition system.

Details about the hardware, the dead-time modelling and the considerations made to undertake those simulations may be found in our previous work [6] along with the literature already mentioned. Results obtained reflect the count rate capability of the given architecture based upon the hardware configuration from geometry to readout electronics. The net gain attainable when compared to a lossless design is computed as well as the Noise Equivalent Count Rate (NECR) to favour the comparison in cases of change applied to the layout.

The NECR is computed from the rates of Trues (T), Scattered (S) and Randoms (R) effectively recorded and used for the images reconstruction according to formula 1.

$$NECR = \frac{T^2}{T + S + 2R} \quad (1)$$

NECR has been introduced many years ago as a standard figure of merit to compare PET performances and yields an assessment of the effective rate at which trues would be collected if noise were in trues alone [3]. Such a formalism is sensitive to adjustments made to the source or the geometry [7] but find its usefulness in predicting how subsequent changes in Trues, Scattered and Randoms do affect the image quality.

Besides the scattered event inherently appearing as a function of the experimental configuration, the random rate is easily predictable in PET. For an individual line of response (LOR), it is given by:

$$R_{rate} = 2 N_a N_b \tau_{coinc} \quad (2)$$

where $N_a$ and $N_b$ are the single event rate for the two crystals defining the LOR and $\tau_{coinc}$ the hardware coincidence timing window width. The total number of random in the image is the sum over all the LORs, keeping overall the linear dependence on $\tau_{coinc}$ [7]. Singles of which energy agrees with the expectations involve the electronics to trigger and then recording of the hit. Again the rate depends on the layout and following the two counterparts energies, they are regarded either as trues or scattered at this stage. Part of the signal is purely noise from randoms which naturally outline the importance of the coincidence window time setting. An assessment of the random rate may be performed, but after the fact only. The latter aspect is none the least, no doubt about the random rate calculation and the subtraction reliability but the noise resulting from the statistical variations in this rate remains [7]. A reduction of the time coincidence window towards the limit will find its usefulness in further veto events prior recording, increasing by the way the NECR. It will only be possible through an improvement of the timing resolution.

## 3. Timing resolution

In state of the art tomographs with fast scintillators, a minimum coincidence timing resolution of 2-3ns fwhm can be achieved [7]. In [1] and [8] the authors pointed out the components that limit the overall timing resolution in the Accel, associating each to a contribution as listed in table 1. The experimental setup is given in figure 1.

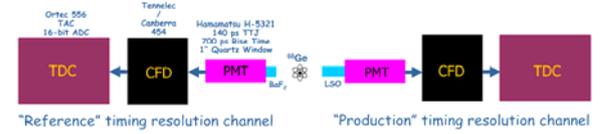

Fig. 1. Setup used to determine the time contributions.

Table 1. Individual contributions to the timing resolution

| Component | Contribution (fwhm) |
|---|---|
| LSO (3x3x30mm$^2$) | 336ps |
| Light sharing (block) | 454ps |
| PMT | 422ps |
| PMT array | 274ps |
| CFD | 1354ps |
| TDC | 2000ps |

The first channel was left unchanged along the study and consisted in a sequence of very high-end electronics parts which serves as reference. The crystal used here is BaF$_2$, well designed for timing. On the right hand side a similar chain appears, in which the components were replaced in turn by their production counterpart used in the commercial camera. Value for the LSO contribution was drawn from [8] so as to better fit with the PET crystals size requirement. It clearly appears that the CFD and TDC bring the main contributions and hence limit the minimum resolution achievable. To be noticed, the value for the CFD is a "raw" value and would have to be increased to account for the shaping that will be used in practice.

The worst case scenario involving a positron annihilation at the far end of a patient while into contact with the instrument gantry as shown in figure 2 imposes a minimum coincidence windows of

$$\tau_{coinc_{min}} = \frac{D_{max}}{c} + \sqrt{2} \times \sqrt{\Sigma (contributions)^2} \quad (3)$$

required for the photons to flight across the detector.

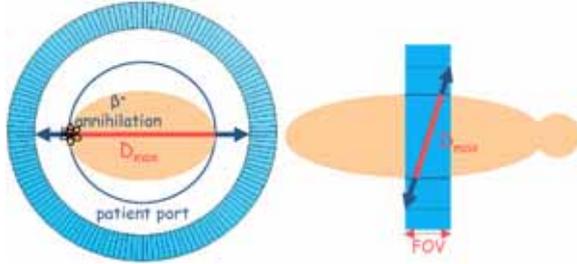

Fig. 2. Sketch of the least favourable location of emission

A numerical application of equation 3 with the informations provided and $D_{max}$ of about 60cm gives a minimum possible setting for the coincidence window of about 5.5ns (2ns + 3.5ns), which well agrees with the typical value for this machine.

### 4. TOF Implementation

In the concern of pure optimization of the coincidence time window, the latter may not be set smaller than the physical limit of 2ns for a ring of 80cm diameter according to formula 3. This prevents from extra noise variance reduction beyond that point due to the number of detected randoms in non-TOF PET. In the latter indeed, the location of the emission point along the LOR is left unknown by conception, so the algorithm used to reconstruct the image uniformly increments all the pixels of the chord.

A TOF solution proposes to measure the difference between times arrivals of back-to-back photons with high resolution to compute somehow the "depth" of interaction along the LOR of concern. This leads in a kind of 3-dimensional localization of the positron position of which accuracy may be measured using

$$\Delta x = \frac{c}{2} \times \Delta t \quad (4)$$

with $\Delta x$ the error committed on the position, $c$ the speed of light and $\Delta T$ the coincidence timing resolution of the scanner. With a $\Delta T$ of 500ps the position location would be constrained to a line segment of about 7.5cm. Even if it was realized that such depth accuracy did not yield in any spatial resolution improvement, it reduces the statistical noise in the reconstructed image as long as the emission source is longer in size than the line segment so regarded [2] [7]. An expression for the noise variance reduction is given by equation 5

$$f = \frac{D}{\Delta x} = \frac{2D}{c\Delta t} \quad (5)$$

with $D$ the object diameter. The reduction in variance does not concern trues only but also scattered and randoms as shown in [9] and [10]. This important consideration involves that the noise added by randoms continues to diminish with the timing resolution enhancement, yet the hardware coincidence window remains at the 2ns bounded by the geometry.

In [7], "For random and scattered events, the effective diameter of the emission source (i.e., the diameter of the object that would be reconstructed using just the random or scattered events) is larger than the actual emission source and can be approximated by the camera's patient port diameter." So regarded and recalling equation 5, the efficiency gain from scattered and randoms are expected to outclass that of trues, hence assuming a global effective source diameter of size that of the phantom (patient), the possible noise variance improvement in the final image will be underestimated.

### 5. Reducing dead-time and noise through a novel electronics scheme

A solution for dead-times to be lowered at bocks and buckets would be to pixelise the detector. A minimum integration time would then be allowed at blocks while leaving the surrounding crystals active. A more powerful solution would be to use a free-running electronics that constantly integrates the incoming signals and clears out count losses that might occur. Merging the two ideas, the light sharing and PMT array contributions to the time resolution would also be suppressed.

The number of channels imposed by the pixelation of the ring tends towards ASICs for reasons of cost and space. However, while very high-end TDC have successfully been integrated in the past years (not compromise-free on complexity, consumption, die surface area and noise), there is still a lack of on-chip CFD at the present time as difficulties arise when delay lines are needed. Both analogue derivation of the signal and shared-constant networks have been imagined to make up for that issue, but electrical noise seems the major trade-off when very small timing resolution is aimed.

Successive events passing through the very front-end electronics freely would require to be discriminated one from the others in energy and time at some point. The originality of our design resides in suppressing the need of dedicated timing channel and the presence of CFD and TDC that comes with. A conversion of the shaped signal to digital applies and information about the energy time are recovered from samples using an optimal filtering method similar to those used in HEP. This well fulfils the integration requirement whereas it makes possible the pile-up suppression and removing of the CFD and TDC contributions to the timing resolution by the way.

Overall, such a scheme would suppress all the dead-times with respect to a conventional chain, optimize the timing resolution by keeping only the very necessary contributions and virtually leave no limitation after digitization about the algorithm to be implemented, including TOF.

In practice, we followed our strategy and pitched on a free running pipelined architecture of which a synoptic is shown in figure 3. So far, the very front-end electronics including the charge preamplifier and the differential CR-RC shaper has been processed and is being tested. The design has been optimized to handle a charge of 50fC from an Hamamatsu APD. A view of the layout of this very front-end electronics for one channel is presented in figure 4.

Concerning the sampling differential flash ADC, the design is ongoing, with a 7 bits encoding at 50MHz. These properties have been laid down to suit the optimal filtering needs for a 500ps coincidence timing resolution. Only will remain the trigger logic, supposed to proceed to the online selection of the events of interest for later image reconstruction.

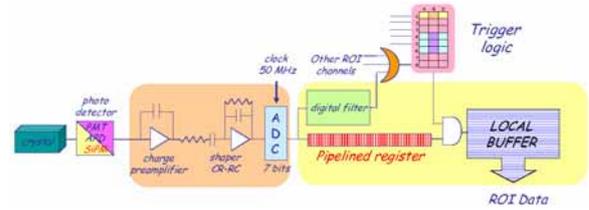

Fig. 3. Synoptic of the lossless and TOF capable electronics proposed.

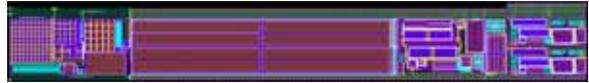

Fig. 4. Layout of the preamplifier and the shaper.

## 6. Signal reconstruction with Optimal filtering

Optimal filtering consists in a weighted sum of signal samples to work out the energy and time while minimizing the noise impact. For this method to be efficient and its implementation worthwhile, the incoming signal must be reproducible in shape, not necessarily in amplitude as a normalization applies. Literature covers this subject in details [11] [12].

Measurements we carried out for LSO and LYSO showed that the naturally occurring signal shape was near constant whereas no shaping was employed at first. This tendency was further noticed with shaping as expected, shaping in any way necessary to remove the electrical noise and slow the signal at ADC input.

Optimal filtering adapted to our particular case yields the results listed in the table 2. Figures express one channel timing resolutions in nanoseconds (sigma). As suspected, a compromise between the sampling frequency and the number of bits is to be accepted. Simulations have shown that a 7 bits ADC may play the role with satisfaction with a frequency of 37MHz or over. A solution involving a 8 bits ADC would do the same at a frequency close to 20MHz only, but would require twice the number of comparators. For this reason, we preferred to stick to the 7 bits architecture.

All the results summarized in this table represent the mathematical limits computed for an ideal signal free of electrical noise. It will not possibly be transcended and the need to compensate the potentially added noise appears mandatory. This bring us to the 50MHz 7 bits ADC already mentioned in section 5. The related error distribution gives a timing accuracy of 300ps fwhm, or 430ps in coincidence.

Table 2. Optimal filtering technique applied to PET signals

| Samples | 8 | 12 | 16 | 20 | 24 | 28 |
|---|---|---|---|---|---|---|
| F/MHz | 24.4 | 36.6 | 48.8 | 61 | 73.2 | 85.4 |
| N bits | | | | | | |
| 4 | 1.473 | 1.213 | 0.993 | 1.046 | 0.890 | 0.699 |
| 5 | 0.743 | 0.585 | 0.513 | 0.456 | 0.419 | 0.371 |
| 6 | 0.366 | 0.298 | 0.259 | 0.229 | 0.206 | 0.194 |
| 7 | 0.184 | 0.152 | 0.129 | 0.115 | 0.105 | 0.096 |
| 8 | 0.094 | 0.082 | 0.065 | 0.058 | 0.052 | 0.049 |
| 9 | 0.05 | 0.052 | 0.034 | 0.030 | 0.027 | 0.025 |
| 10 | 0.03 | 0.041 | 0.02 | 0.018 | 0.014 | 0.013 |

## 7. Results and discussion

Simulations were performed for a hypothetical cylindrical phantom of human size (175cm long, 40cm diameter). As we hinted at, the camera basis is the HR+, updated to better fit with the present situation and come closer to the Accel. It involved the replacement of the BGO by LSO, a modification of the processing times at blocks and buckets, and the narrowing of the coincidence window down to 6ns. The activity considered was of 5.5kBq/cc. A first simulation showed a 17% saturation of the electronics with respect to a lossless design, which leaves only a 20% room for improvement on count-rate. With the latter design and according to what has been developed in section 3, only the scintillator and photo-detector contributions to the timing resolution would remain. Assuming an APD as the photo-detector with its typical 50-100ps timing resolution, it would not penalize the LSO characteristics and make the 500ps achievable. In these conditions, the coincidence window width might be set at 2.5ns (2ns + 0.5ns). A Related simulation showed that the NECR goes up by a factor 1.7, to be weighted by the possible 1.2 gain from electronics. It leads overall to a net improvement of the NECR by a factor 2. With such a configuration, the TOF algorithm will naturally be included and we expect a noise variance enhancement of about 5 according to formula 5.

## 8. Conclusion

It has been known for long that enhancing the timing resolution of the detection chain in PET scanners significantly improves their events collection efficiency through a possible decrease of the coincidence window width. A TOF implementation begins to be worthwhile when the coincidence timing resolution drops below the nanosecond by bringing an information about the depth of interaction along the LORs. Besides this, the electronics schemes actually employed still display dead-times that make them to saturate. Merging both ideas, we proposed a new electronics synoptic well inspired of HEP experiments that appears inherently a high-end candidate both for count-rate and timing. A TOF implementation would be considered as a resolution of 500ps may be achieve via optimal filtering, leading in an awaited noise variance improvement of 5 in the final images. Concerning, the NECR, the net gain expected is of about 2.